\documentclass[prl,twocolumn,superscriptaddress]{revtex4-2}
\usepackage{graphicx,color}
\usepackage{type1cm}
\usepackage[usenames,svgnames]{xcolor}
\usepackage{url}
\usepackage{natbib}
\usepackage[colorlinks=true,linkcolor=blue,citecolor=blue,breaklinks=true]{hyperref}

\usepackage{amsmath,amssymb,bm,amsthm}
\usepackage{newtxtext,newtxmath}
\usepackage{braket}

\newcommand{\sectionprl}[1]{{\par\it #1.---}}

\begin{document}
\nocite{apsrev41control}


\title{Rigorous Bounds on the Heating Rate in Thue-Morse Quasiperiodically and Randomly Driven Quantum Many-Body Systems}

\author{Takashi Mori}
\affiliation{RIKEN Center for Emergent Matter Science (CEMS), Wako 351-0198, Japan}
\author{Hongzheng Zhao}
\affiliation{Blackett Laboratory, Imperial College London, London SW7 2AZ, United Kingdom}
\author{Florian Mintert}
\affiliation{Blackett Laboratory, Imperial College London, London SW7 2AZ, United Kingdom}
\author{Johannes Knolle}
\affiliation{Department of Physics TQM, Technische Universit\"at M\"unchen, James-Franck-Stra\ss e 1, D-85748 Garching, Germany}
\affiliation{Munich Center for Quantum Science and Technology (MCQST), 80799 Munich, Germany}
\affiliation{Blackett Laboratory, Imperial College London, London SW7 2AZ, United Kingdom}

\author{Roderich Moessner}
\affiliation{Max-Planck-Institut f\"ur Physik komplexer Systeme, N\"othnitzer Stra\ss e 38, 01187 Dresden, Germany}

\begin{abstract}
The nonequilibrium quantum dynamics
of closed many-body systems is a rich yet challenging field. While recent progress for periodically driven (Floquet) systems has yielded a number of rigorous results, our understanding on quantum many-body systems driven by rapidly varying but a- and quasi-periodic driving is still limited.
Here, we derive rigorous, non-perturbative, bounds on the heating rate in quantum many-body systems under Thue-Morse quasi-periodic driving and under random multipolar driving, the latter being a tunably randomized variant of the former. In the process, we 
derive a static effective  Hamiltonian that describes the transient prethermal state, including the dynamics of local observables. 
Our bound for Thue-Morse quasi-periodic driving suggests that the heating time scales like $(\omega/g)^{-C\ln(\omega/g)}$ with a positive constant $C$ and a typical energy scale $g$ of the Hamiltonian, in agreement with our numerical simulations. 
\end{abstract}
\maketitle


\begin{acknowledgments}
\end{acknowledgments}

\sectionprl{Introduction}
When a thermally isolated ergodic quantum many-body system is subjected to time-dependent external fields, it eventually enters an entirely featureless (`infinite-temperature') state.
The route it takes involves intriguing non-equilibrium dynamics. The technical challenges involved in describing non-equilibrium many-body dynamics such as this are formidable, and specifically rigorous results are very rare.

Periodically driven (Floquet) systems provide a notable exception: here  the heating rate is exponentially small at high $\omega$~\cite{Kuwahara2016,Mori2016b,Abanin2017,Abanin2017a}.
This result implies that a Floquet system generically exhibits prethermalization behavior~\cite{Mori2018_review}: it first relaxes to a long-lived prethermal state, followed by slow heating to infinite temperature.
The prethermal state is described by a static effective Hamiltonian obtained by the high-frequency (Magnus) expansion.
This opens the possibility of Floquet engineering, i.e.,  implementing a desired effective Hamiltonian by applying periodic driving~\cite{Eckardt2017, Oka2019}.
Floquet prethermal states can even realize novel phases of matter such as prethermal Floquet time crystals in clean systems~\cite{Else2017}.
In experiments, Floquet prethermalization has been observed for ultra-cold atoms in a driven optical lattice~\cite{Rubio-Abadal2020} and for nuclear spins in Fluorapatite~\cite{Peng2021}.

Recently, the focus has shifted beyond fully periodic drives to \textit{quasi-periodic driving}~\cite{Verdeny2016, Nandy2017, Dumitrescu2018, Giergiel2019, Ray2019, Zhao2020, Else2020, Mukherjee2020}. A further step was taken by considering \textit{structured random driving}~\cite{Zhao2020}, which permits  interpolating between fully random and quasi-periodic drives. 
Under fast random driving, heating is often swift, and no prethermalization is observed.
However, recent reports suggest that prethermalization can still be encountered under fast quasi-periodic driving~\cite{Dumitrescu2018, Zhao2020} or certain structured random drives~\cite{Zhao2020}. For the former, rigorous results on the heating rates are available for {\it continuously} varying quasi-periodic driving~\cite{Else2020}.
However, no such results exist for discrete quasi-periodic driving, such as Fibonacci~\cite{Dumitrescu2018, Ray2019} and Thue-Morse~\cite{Nandy2017, Zhao2020, Mukherjee2020} driving.

Here, we provide rigorous bounds on the heating rate for Thue-Morse quasi-periodic driving, as well as for the random multipolar driving introduced in Ref.~\cite{Zhao2020}. These turn out to differ from those of the abovementioned settings: we find a power-law bound with a tunable exponent in random multipolar driving, as well as a rate vanishing faster than any power for Thue-Morse.
In a Magnus expansion, the heating bounds are derived alongside a static effective Hamiltonian that describes prethermal states and their transient nonequilibrium dynamics. These results are non-perturbative in driving amplitude, and  they provide a transparent connection to the lore of Floquet systems.

In the following, we first introduce the class of models under consideration, and the Magnus expansion used to derive our core results. These we supplement by numerical investigations of actual model Hamiltonians. We describe possible generalisations and close with a discussion and outlook. 

\sectionprl{Setting}
We consider two Hamiltonians $H_+$ and $H_-$ for a lattice system.
The time evolution operators over a time period $T$ generated by $H_\pm$ are denoted by $U^{(\pm)}=e^{-iH_\pm T}$.
The quantity $\omega=2\pi/T$ is referred to as the ``frequency'' of the drive.
For a periodic sequence of $U_\pm$, Floquet prethermalizationn occurs and the lifetime of a prethermal state scales exponentially in $\omega$.
On the other hand, for random sequences of $U_\pm$, the heating time remains finite even for $\omega\rightarrow\infty$, and hence no prethermalization occurs in general.

Random multipolar driving~\cite{Zhao2020} ($n$-RMD in short) generated by $U_\pm$ involves a random sequence of ``dipoles'' $U_1^{(\pm)}$, where $U_1^{(+)}=U_-U_+$ and $U_1^{(-)}=U_+U_-$, or ``quadrupoles''  $U_2^{(\pm)}$, where $U_2^{(+)}=U_+U_-U_-U_+$ and $U_2^{(-)}=U_-U_+U_+U_-$, and so on recursively:
$n$-multipoles $U_n^{(\pm)}$ are given by $U_n^{(+)}=U_{n-1}^{(-)}U_{n-1}^{(+)}$ and $U_n^{(-)}=U_{n-1}^{(+)}U_{n-1}^{(-)}$ starting from $U_0^{(\pm)}=U_\pm$.
In Ref.~\cite{Zhao2020}, it is found that the heating time scales like $\omega^{2n+1}$ with $n\geq 1$, which corresponds to a random sequence of $U_n^{(\pm)}$.
The limit of $n\to\infty$ of $n$-RMD corresponds to the Thue-Morse quasi-periodic driving.
Prethermalization also occurs in this limit, but the situation has not been fully clear: the heating time is longer than algebraic in $\omega$ and was found to be consistent with exponential scaling for finite size numerical calculations~\cite{Zhao2020}.

An increasing heating time $\sim\omega^{2n+1}$ with $n$ for $n$-RMD has been derived within the linear-response regime, perturbatively via Fermi's golden rule.
For example, when $n=0$ (a random sequence of $U_\pm$), Fermi's golden rule predicts a heating time  proportional to $\omega$, but it actually remains finite in the limit of $\omega\to\infty$~\footnote{See Supplementary Material for the proof of Eq.~(\ref{eq:Magnus_formula}), the detail of Fermi's golden rule calculations, and the formal expression of Eq.~(\ref{eq:Lemma}) and its proof.}.
In this example, for any fixed driving amplitude, the driving cannot be regarded as perturbatively weak for sufficiently high frequencies which motivates us to derive rigorous results.
As slow heating is numerically observed in the Thue-Morse and random cases even for strong driving, it is any rate imperative to go beyond linear-response.

Below we derive such rigorous upper bounds on the heating rate for $n$-RMD and the Thue-Morse quasi-periodic driving without the assumption of weak driving.
The previous results for periodic driving~\cite{Mori2016b} can be discussed in a parallel way.

\sectionprl{Magnus expansion}
Since $U_n^{(\pm)}$ is a sequence of $U_+$ and $U_-$ of length $2^L$, it represents unitary time evolution over a time $2^nT$.
Let us express $U_n^{(\pm)}$ in the form $U_n^{(\pm)}=e^{-i\tilde{H}_n^{(\pm)}2^nT}$, where $\tilde{H}_n^{(\pm)}$ is a certain Hermitian operator.
It is in general hopeless to obtain an exact explicit expression of $\tilde{H}_n^{(\pm)}$ for many-body systems, but its high-frequency expansion is available.
The Magnus expansion of $\tilde{H}_n^{(\pm)}$ is given as follows:
\begin{equation}
\tilde{H}_n^{(\pm)}=\sum_{m=0}^\infty (2^nT)^m\Omega_{n,m}^{(\pm)}.
\end{equation}
By introducing an instantaneous Hamiltonian $H_n^{(\pm)}(t)$, which is either $H_+$ or $H_-$ for each time step $t\in [\ell T,(\ell+1)T)$ depending on whether $\ell$th unitary of $U_n^{(\pm)}$ is $U_+$ or $U_-$, $U_n^{(\pm)}$ is expressed as $U_n^{(\pm)}=\mathcal{T}e^{-\int_0^{2^nT}dt\,H_n^{(\pm)}(t)}$, where $\mathcal{T}$ stands for the time-ordering operator.
It should be noted that the lowest term of the Magnus expansion is given by the time-averaged Hamiltonian:
\begin{equation}
\Omega_{n,0}^{(\pm)}=\frac{1}{2^nT}\int_0^{2^nT}dt\,H_n^{(\pm)}(t)=\frac{H_++H_-}{2}.
\end{equation}
Explicit expressions of higher order terms are provided in, e.g., Refs.~\cite{Kuwahara2016, Mori2016b}.
By using the relations $U_n^{(\pm)}=U_{n-1}^{(\mp)}U_{n-1}^{(\pm)}$, we can obtain $\Omega_{n,m}^{(\pm)}$ recursively starting from $n=0$: $\Omega_{0,m}^{(\pm)}=\delta_{m,0}H_\pm$.

A crucial observation here is that the Magnus expansion in $U_n^{(\pm)}$ has the following remarkable property: for each $n$,
\begin{equation}
\Omega_{n,m}^{(+)}=\Omega_{n,m}^{(-)}=2^{-m(n-m)}\frac{\Omega_{m,m}^{(+)}+\Omega_{m,m}^{(-)}}{2}\quad\text{for all }m\leq n-1.
\label{eq:Magnus_formula}
\end{equation}
This property stems from self-similarity of the Thue-Morse sequence, and it plays an essential role in deriving rigorous bounds on the heating rate discussed below.
We can prove Eq.~(\ref{eq:Magnus_formula}) by induction, see Supplementary Material for  details~\footnotemark[1].

In this way, although $\tilde{H}_n^{(+)}$ differs from $\tilde{H}_n^{(-)}$, their Magnus expansions coincide with each other up to $(n-1)$th order.
Therefore, $H_\mathrm{eff}^{(n)}$ plays the role of a static effective Hamiltonian for time evolution generated by an arbitrary sequence of $U_n^{(+)}$ and $U_n^{(-)}$.
It should be noted that by using Eq.~(\ref{eq:Magnus_formula}), $H_\mathrm{eff}^{(n)}$ is also expressed as
\begin{equation}
H_\mathrm{eff}^{(n)}=\sum_{m=0}^{n-1}(2^mT)^m\frac{\Omega_{m,m}^{(+)}+\Omega_{m,m}^{(-)}}{2}.
\end{equation}
Thus the $m$th order term of the Magnus expansion of $\tilde{H}_n^{(\pm)}$ is independent of $n$ as long as $n\geq m$, which enables us to define the high-frequency expansion of the effective Hamiltonian for the Thue-Morse quasi-periodic driving $n\to\infty$.

\sectionprl{Rigorous bounds on heating rate}
Time evolution under a generic time-dependent Hamiltonian $H(t)$ from time $t=0$ to $t=\tau$ is expressed by a unitary operator $U_\tau=\mathcal{T}e^{-i\int_0^\tau dt\,H(t)}=:e^{-i\tilde{H}\tau}$.
Let us consider the Magnus expansion of $\tilde{H}$: $\tilde{H}=\sum_{m=0}^\infty\tau^m\Omega_m$.
Its truncation at $n$th order is denoted by $\tilde{H}^{(n)}=\sum_{m=0}^{n}\tau^m\Omega_m$.
Under this general setting, the following inequality is proved for a generic local Hamiltonian~\footnote{Here, ``local'' Hamiltonian implies a $k$-local Hamiltonian, which means that the Hamiltonian contains up to $k$-site interactions. The interaction range may be arbitrarily long.}:
\begin{equation}
\|U_\tau^\dagger\tilde{H}^{(n)}U_\tau-\tilde{H}^{(n)}\|\lesssim (n+1)!(g\tau)^{n+2}gN,
\label{eq:Lemma}
\end{equation}
where $g$ denotes the maximum energy of a single site and $N$ denotes the number of lattice sites (system size).
See Supplementary Material for an exact formal expression of the inequality~(\ref{eq:Lemma}) and its proof~\footnotemark[1].

Exponentially slow heating in Floquet systems is derived by using the inequality~(\ref{eq:Lemma}).
In the case of periodic driving, we choose $\tau$ to be the period $T=2\pi/\omega$ of driving.
In this case $\tilde{H}$ is nothing but the Floquet Hamiltonian $H_F$, which is independent of time.
Therefore, after $M$ periods $(t=MT)$, the possible change of $H_F^{(n)}$ is bounded by
\begin{align}
\frac{\|U_T^{\dagger M}H_F^{(n)}U_T^M-H_F^{(n)}\|}{N}\leq M\frac{\|U_T^\dagger H_F^{(n)}U_T-H_F^{(n)}\|}{N}
\nonumber \\
\lesssim(n+1)!(gT)^{n+2}gM=(n+1)!(Tg)^{n+1}t.
\end{align}
The heating rate $\kappa$ is therefore bounded as $\kappa\lesssim (n+1)!(Tg)^{n+1}$.
Since $n$ is arbitrary, we choose it so that the upper bound of $\kappa$ becomes minimum.
We then have the optimal truncation order $n^*\sim\omega/g$ and $\kappa\lesssim e^{-O(\omega/g)}$.
The heating rate in Floquet systems is thus exponentially small in $\omega$~\cite{Kuwahara2016, Mori2016b}.

In the case of $n$-RMD, we choose $\tau=2^nT$.
Then, $U_\tau$ is either $U_n^{(+)}$ or $U_n^{(-)}$.
As we have already seen, $\tilde{H}^{(n-1)}=H_\mathrm{eff}^{(n)}$ for both $U_n^{(+)}$ and $U_n^{(-)}$.
Thus we have from Eq.~(\ref{eq:Lemma})
\begin{equation}
\|U_n^{(\pm)\dagger}H_\mathrm{eff}^{(n)}U_n^{(\pm)}-H_\mathrm{eff}^{(n)}\|\lesssim n!(2^ngT)^{n+1}gN.
\end{equation}
The time evolution $U_t$ over time $t=M2^nT$ is generated by a certain sequence of $U_n^{(\pm)}$, i.e., $U_t=U_n^{(\sigma_M)}U_n^{(\sigma_{M-1})}\dots U_n^{(\sigma_1)}$ with $\sigma_\ell\in\{+,-\}$.
The change of $H_\mathrm{eff}^{(n)}$ over time $t=M2^nT$ is evaluated as
\begin{align}
\frac{\|U_t^\dagger H_\mathrm{eff}^{(n)}U_t-H_\mathrm{eff}^{(n)}\|}{N}&\leq\sum_{\ell=1}^M\frac{\|U_n^{(\sigma_\ell)\dagger}H_\mathrm{eff}^{(n)}U_n^{(\sigma_\ell)}-H_\mathrm{eff}^{(n)}\|}{N}
\nonumber \\
&\lesssim n!(2^ngT)^nt.
\end{align}
The heating rate $\kappa$ is thus evaluated as 
\begin{equation}
\kappa\lesssim n!(2^ngT)^n\sim n!\left(\frac{2^ng}{\omega}\right)^n.
\label{eq:heating_n}
\end{equation}
This bound shows that heating becomes slower with $n$ for large enough $\omega$.
This upper bound is however not optimal since numerical results indicate $\kappa\sim\omega^{2n+1}$~\cite{Zhao2020}.
Later we will see that the correct scaling of the heating rate is obtained by combining the Magnus expansion for the $n$-RMD developed above with Fermi's golden rule.

The heating rate under the Thue-Morse quasi-periodic driving is also evaluated by using Eq.~(\ref{eq:heating_n}).
Since the Thue-Morse driving is regarded as a certain sequence of $U_n^{(\pm)}$ for arbitrary $n$, the heating rate $\kappa$ in this case satisfies Eq.~(\ref{eq:heating_n}) for arbitrary $n$.
Therefore, the optimal truncation order $n^*$ is determined by minimizing the right-hand-side of Eq.~(\ref{eq:heating_n}).
As a result, we have
\begin{equation}
n^*\sim\ln(\omega/g)
\end{equation}
and
\begin{equation}
\kappa\lesssim e^{-C[\ln(\omega/g)]^2}=\left(\frac{\omega}{g}\right)^{-C\ln(\omega/g)},
\quad C\geq\frac{1}{4\ln 2}.
\label{eq:upper_Thue-Morse}
\end{equation}
This upper bound of $\kappa$ implies that the heating time under the Thue-Morse quasi-periodic driving is \textit{shorter than $e^{O(\omega/g)}$ but longer than any polynomial in $\omega/g$}.

Up to here, we have shown that $H_\mathrm{eff}^{(n)}$ is a quasi-conserved quantity. 
If no (quasi)conserved quantity exists besides $H_\mathrm{eff}^{(n)}$, a prethermal state is described by the Gibbs state $\rho_\mathrm{pre}\propto e^{-\beta_\mathrm{eff}H_\mathrm{eff}^{(n)}}$ with an effective temperature $\beta_\mathrm{eff}^{-1}$, which is determined by the initial value of the expectation value of $H_\mathrm{eff}^{(n)}$.

\sectionprl{Accurate heating rates for $n$-RMD}
For $n$-RMD, we have obtained a rigorous upper bound $\kappa\lesssim O(\omega^{-n})$ for large $\omega$, but it is not optimal as we have already mentioned.
The correct scaling $\kappa\sim\omega^{-(2n+1)}$ is theoretically (but not rigorously) obtained by combining the Magnus expansion technique with Fermi's golden rule.
For $n$-RMD with $n\geq 1$, $\tilde{H}_n^{(\pm)}$ shares the common Magnus expansion $H_\mathrm{eff}^{(n)}$ up to $(n-1)$th order.
Now we investigate the effect of one more higher order term $(2^nT)^n\Omega_{n,n}^{(\pm)}$ by considering it as a perturbation.
Truncated Magnus expansions of $\tilde{H}_n^{(\pm)}$ at $n$th order are given by $\tilde{H}_\mathrm{eff}^{(n,\pm)}:=H_\mathrm{eff}^{(n)}+(2^nT)^n\Omega_{n,n}^{(\pm)}$.
For each time interval $t\in[\ell 2^nT,(\ell+1)2^nT)$, either $\tilde{H}_\mathrm{eff}^{(n,+)}$ or $\tilde{H}_\mathrm{eff}^{(n,-)}$ is chosen randomly.
Therefore the additional term $(2^nT)^n\Omega_{n,n}^{(\pm)}$ is regarded as a random driving with strength proportional to $T^n$.
According to Fermi's golden rule, the heating rate under a random sequence of two Hamiltonians whose difference is proportional to $v$ is given by $\kappa\sim\omega^{-1}v^2$~\footnotemark[1].
In our case $v\sim T^n\sim\omega^{-n}$, and therefore
\begin{equation}
\kappa\sim\omega^{-1}\cdot\omega^{-2n}=\omega^{-(2n+1)},
\label{eq:Fermi_heating}
\end{equation}
which agrees with numerical results~\cite{Zhao2020}.

We emphasize that Fermi's golden rule calculations discussed here are beyond the linear-response argument.
Here we have applied the golden rule to the effective Hamiltonian (random driving of $H_\mathrm{eff}^{(n,\pm)}$), in which nonlinear effects of the driving field is already taken into account.
Indeed, the assumption of weak driving is unnecessary.
Equation~(\ref{eq:Fermi_heating}) is valid as long as the frequency $\omega$ is large enough.

\sectionprl{Comparison with numerical results}
\begin{figure}[t]
\centering
\includegraphics[width=0.48\textwidth]{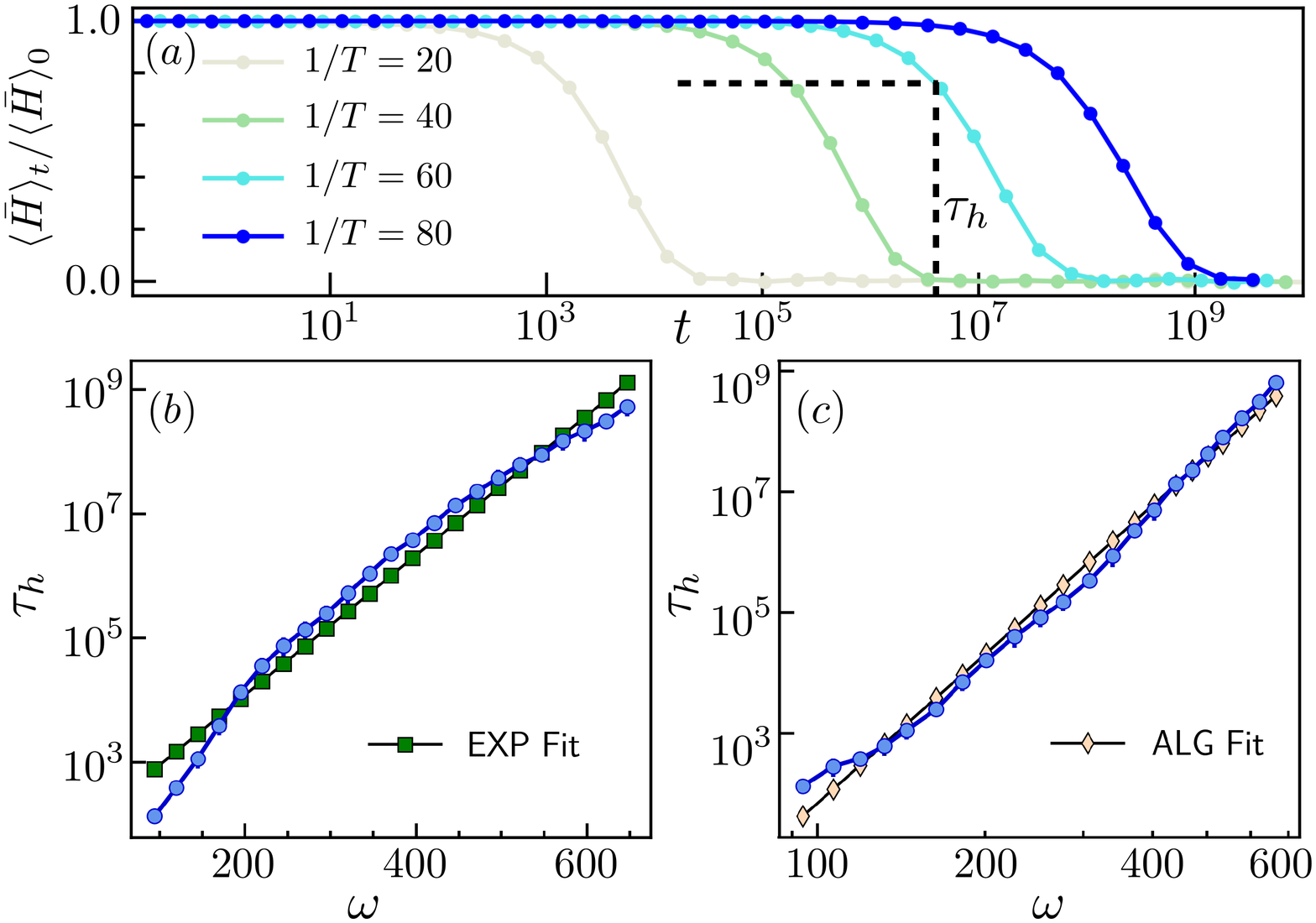}
\caption{(a) Time evolutions of $\braket{\bar{H}}_t/\braket{\bar{H}}_0$ in the model under the Thue-Morse quasi-periodic driving for various driving frequencies. (b) Semi-log plot and (c) log-log plot of the heating rate (blue circles). 
Straight lines in (b) (green squares) and (c) (pink diamonds) correspond to an exponential fit and an algebraic hit, respectively.
In (b), numerical data curves down from an exponential fit.
In (c), numerical data curves up from an algebraic fit.}
\label{fig:fitting}
\end{figure}
We have obtained $\kappa\sim\omega^{-(2n+1)}$ for $n$-RMD by applying Fermi's golden rule to the Magnus expansion, which agrees with numerical results obtained in the previous work~\cite{Zhao2020}.
On the other hand, for the Thue-Morse quasi-periodic driving, we have obtained the not so familiar scaling $\kappa\lesssim (\omega/g)^{-C\ln(\omega/g)}$.
Previous work~\cite{Zhao2020} reported that the heating time looks exponential in $\omega$, while our upper bound (\ref{eq:upper_Thue-Morse}) suggests neither exponential nor algebraic dependence of the heating rate on the frequency.
Although our upper bound does not contradict an exponential dependence, we should carefully analyze numerical data because it is sometimes a tricky problem to determine even whether a given numerical data shows an exponential or algebraic dependence.

In numerical calculations, the following Hamilltonians are considered:
\begin{equation}
H_\pm=\sum_{i=1}^N\left[\sigma_i^z\sigma_{i+1}^z+J_x\sigma_i^x\sigma_{i+1}^x+(B_0\pm B_x)\sigma_i^x+B_z\sigma_i^z\right],
\end{equation}
where $\sigma_i^x,\sigma_i^y,\sigma_i^z$ are Pauli matrices and periodic boundary conditions are imposed.
In this Letter we fix $J_x=0.72$, $B_z=0.49$, $B_x=0.61$, and $B_0=0.21$.
For the Thue-Morse quasi-periodic driving, we calculate the time evolution of the averaged Hamiltonian $\bar{H}=(H_++H_-)/2$, which is regarded as the energy of the system.
Figure~\ref{fig:fitting} (a) shows time evolutions of $\braket{\bar{H}}_t/\braket{\bar{H}}_0$ for various driving frequencies.
Prethermalization occurs and the heating time increases with $\omega\sim 1/T$.
Numerically, the heating time $\tau_h$ is extracted by averaging the time such that $\braket{\bar{H}}_t/\braket{\bar{H}}_0=0.75, 0.75\pm 0.03, 0.75\pm 0.06$, where $\braket{\bar{H}}_t$ denotes the expectation value of the energy at time $t$.
The heating rate $\kappa$ is just given by $\kappa=\tau_h^{-1}$.

\begin{figure}[t]
\centering
\includegraphics[width=0.3\textwidth]{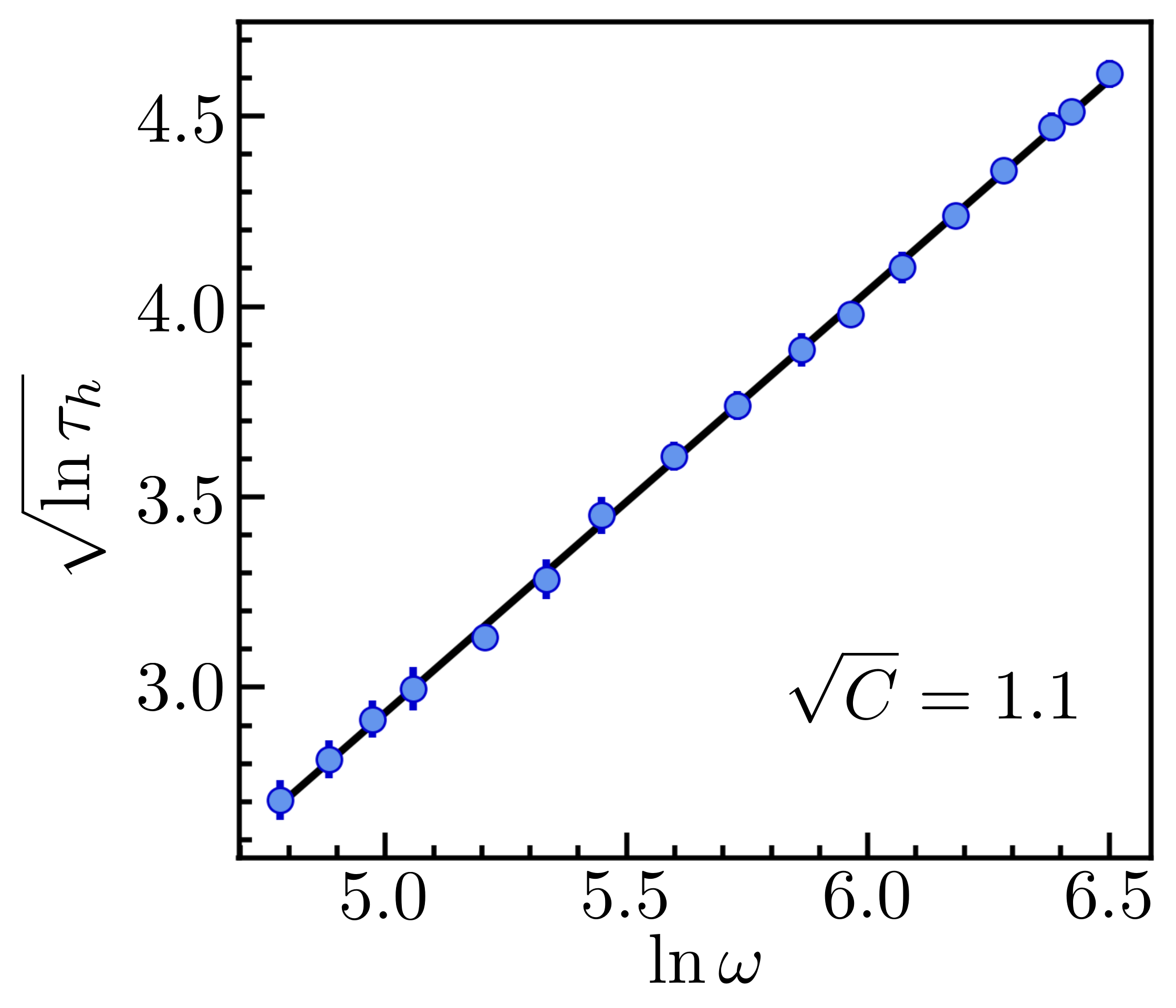}
\caption{Plot of $\sqrt{\ln\tau_h}$ against $\ln\omega$.
Numerical data (blue circles) fits well with a straight line of slope $\sqrt{C}=1.1$, which implies that $\tau_h$ behaves like $\tau_h\sim e^{C[\ln(\omega/g)]^2}$, as suggested by Eq.~(\ref{eq:upper_Thue-Morse}).}
\label{fig:scaling}
\end{figure}

Figure~\ref{fig:fitting} (b) and (c) show a semi-log plot and a log-log plot of the heating time $\tau_h$, respectively.
In the semi-log (log-log) plot, numerical data curves down (up) from a straight line, which indicates that the heating time is shorter than exponential but longer than algebraic in $\omega$.
This is consistent with the upper bound given in Eq.~(\ref{eq:upper_Thue-Morse}).
In Fig.~\ref{fig:scaling}, we plot the function $\sqrt{\ln\tau_h}$ against $\ln\omega$.
If $\kappa\approx e^{-C[\ln(\omega/g)]^2}$ as suggested by Eq.~(\ref{eq:upper_Thue-Morse}), we will have a straight line of slope $\sqrt{C}$.
Numerical data fits well with a straight line of slope $\sqrt{C}=1.1$.
Although the constant $C\geq 1/(4\ln 2)$ in our upper bound~(\ref{eq:upper_Thue-Morse}) seems not tight, our upper bound captures a correct frequency dependence of the heating rate.

\sectionprl{Transient dynamics}
Finally, we make a remark that the effective Hamiltonian $H_\mathrm{eff}^{(n)}$ not only describes a prethermal state, but also governs transient dynamics of local observables under an arbitrary sequence of $U_n^{(\pm)}$.
Let us assume that the Hamiltonians $H_\pm(t)$ describe a $d$-dimensional lattice system with short-ranged interactions (interactions decay exponentially fast).
In Ref.~\cite{Mori2018}, the following inequality has been proved for any local operator $O_X$ that acts nontrivially only on a set of sites $X\subset\{1,2,\dots,N\}$ and is normalized as $\|O_X\|=1$:
\begin{equation}
\left\|U_t^\dagger O_XU_t-e^{iH_\mathrm{eff}^{(n)}t}O_Xe^{-iH_\mathrm{eff}^{(n)}t}\right\|\lesssim |X|n!(2^ngT)^nt^{d+1},
\label{eq:dynamics}
\end{equation}
where $t=M2^nT$ and $U_t=U_n^{(\sigma_M)}U_n^{(\sigma_{M-1})}\dots U_n^{(\sigma_1)}$ with an arbitrary sequence of $\sigma_\ell\in\{+,-\}$.
This inequality is proved by using the inequality like Eq.~(\ref{eq:Lemma}) and the Lieb-Robinson bound~\cite{Lieb1972, Bravyi2006}.

The inequality~(\ref{eq:dynamics}) tells us that the effective Hamiltonian approximates the dynamics of local observables during a time $\tau_d$, where $\tau_d\gtrsim(\omega/g)^{n/(d+1)}$ for $n$-RMD and $\tau_d\gtrsim \exp\left\{\frac{C}{d+1}[\ln(\omega/g)]^2\right\}$ for the Thue-Morse quasi-periodic driving.

\sectionprl{Some extensions}
Our derivation essentially relies on Eq.~(\ref{eq:Magnus_formula}), which is tailored to the Thue-Morse sequence.
Our results allow some extensions.
First, we have assumed static $H_\pm$, but our main results, i.e. Eqs.~(\ref{eq:heating_n}), (\ref{eq:upper_Thue-Morse}), (\ref{eq:Fermi_heating}), and (\ref{eq:dynamics}), hold for time-dependent $H_\pm(t)$ also, because Eq.~(\ref{eq:Magnus_formula}) is not affected by this modification.
Second, the Thue-Morse sequence can be generalised.
For example, consider three unitaries $U_0=e^{-iH_0T}$, $U_1=e^{-iH_1T}$, and $U_2=e^{-iH_2T}$, where $H_\sigma$ ($\sigma=0,1,2$) are local Hamiltonians.
We recursively define $U_n^{(\sigma)}$ as $U_n^{(\sigma)}=U_{n-1}^{(\sigma+2)}U_{n-1}^{(\sigma+1)}U_{n-1}^{(\sigma)}$, where the sums $\sigma+1$ and $\sigma+2$ are defined modulo 3.
Then, a similar upper bound for a random sequence of $\{U_n^{(\sigma)}\}$ or for a quasi-periodic sequence defined by $\lim_{n\to\infty}U_n^{(\sigma)}$.
However, Eq.~(\ref{eq:Magnus_formula}) is in general not satisfied for other quasi-periodic driving such as the Fibonacci driving.
It is an open problem to rigorously evaluate the heating rate for general quasi-periodic drives.

\sectionprl{Conclusion}
We have derived rigorous upper bounds on the heating rate for quantum many-body systems under quasi- and a-periodic driving with tunable structured randomness. These differ from those for  Floquet~\cite{Kuwahara2016, Mori2016b, Abanin2017, Abanin2017a} and smooth quasi-periodic  driving~\cite{Else2020}.

We emphasize that nonperturbative and rigorous results for aperiodic -- tunable yet random -- driven systems are very rare.
Remarkably, such rigorous results are established based on mathematical machineries developed for regular systems (i.e. periodically driven systems)~\cite{Kuwahara2016, Mori2016b}.
It is an interesting problem to investigate whether there are any other structures to the driving which give rise to systematically understandable behavior.



\begin{acknowledgements}
This work was in part supported by Japan Society for the Promotion of Science KAKENHI Grants No. 19K14622, by the Deutsche Forschungsgemeinschaft  under  cluster of excellence ct.qmat (EXC 2147, project-id 390858490), and by a Doctoral-Program of the German Academic Exchange Service (DAAD) Fellowship. We acknowledge support from the Imperial-TUM flagship partnership.
\end{acknowledgements}

\bibliography{apsrevcontrol,physics}

\end{document}


\nocite{apsrev41control}

\title{Supplementary Material}

\author{Takashi Mori}
\affiliation{RIKEN Center for Emergent Matter Science (CEMS), Wako 351-0198, Japan}
\author{Hongzheng Zhao}
\affiliation{Blackett Laboratory, Imperial College London, London SW7 2AZ, United Kingdom}
\author{Florian Mintert}
\affiliation{Blackett Laboratory, Imperial College London, London SW7 2AZ, United Kingdom}
\author{Johannes Knolle}
\affiliation{Department of Physics TQM, Technische Universit\"at M\"unchen, James-Franck-Stra\ss e 1, D-85748 Garching, Germany}
\affiliation{Munich Center for Quantum Science and Technology (MCQST), 80799 Munich, Germany}
\affiliation{Blackett Laboratory, Imperial College London, London SW7 2AZ, United Kingdom}

\author{Roderich Moessner}
\affiliation{Max-Planck-Institut f\"ur Physik komplexer Systeme, N\"othnitzer Stra\ss e 38, 01187 Dresden, Germany}

\maketitle

\setcounter{equation}{0}
\def\theequation{S\arabic{equation}}
\setcounter{figure}{0}
\def\thefigure{S\arabic{figure}}
\setcounter{section}{0}
\def\thesection{\Alph{section}}
\renewcommand*{\citenumfont}[1]{S#1}
\renewcommand*{\bibnumfmt}[1]{[S#1]}

\section{Derivation of Eq. (3)}
Let us prove Eq. (3) in the main text, i.e., for each $n$,
\begin{equation}
\Omega_{n,m}^{(+)}=\Omega_{n,m}^{(-)}=2^{-m(n-m)}\frac{\Omega_{m,m}^{(+)}+\Omega_{m,m}^{(-)}}{2}\quad\text{for all }m\leq n-1.
\label{eq:Magnus_formula}
\end{equation}
First, we assume Eq.~(\ref{eq:Magnus_formula}) for a certain $n$ [it is readily checked that Eq.~(\ref{eq:Magnus_formula}) holds for $n=1$].
Let us define
\begin{equation}
H_\mathrm{eff}^{(n)}=\sum_{m=0}^{n-1}(2^nT)^m\Omega_{n,m}^{(+)}=\sum_{m=0}^{n-1}(2^nT)^m\Omega_{n,m}^{(-)}=\sum_{m=0}^{n-1}(2^mT)^m\frac{\Omega_{m,m}^{(+)}+\Omega_{m,m}^{(-)}}{2},
\end{equation}
which corresponds to a truncation of the Magnus expansion at $(n-1)$th order.
Then, $U_{n+1}^{(\pm)}=U_n^{(\mp)}U_n^{(\pm)}$ are given by
\begin{align}
U_{n+1}^{(\pm)}=\exp\left[-i\left(H_\mathrm{eff}^{(n)}2^nT+\sum_{m=n}^\infty\Omega_{n,m}^{(\mp)}(2^nT)^{m+1}\right)\right]
\nonumber \\
\times\exp\left[-i\left(H_\mathrm{eff}^{(n)}2^nT+\sum_{m=n}^\infty\Omega_{n,m}^{(\pm)}(2^nT)^{m+1}\right)\right].
\end{align}
By using the Baker-Campbell-Hausdorff formula, this is expressed by a single exponential.
Although $e^Ae^B\neq e^{A+B}$ when $[A,B]\neq 0$, correction terms due to non-commutability in our case do not appear up to the order of $T^{n+1}$ [in our case, $[A,B]=O(T^{n+2})$].
We therefore obtain
\begin{align}
U_{n+1}^{(\pm)}=e^{-i\left[H_\mathrm{eff}^{(n)}2^{n+1}T
+\left(\Omega_{n,n}^{(+)}+\Omega_{n,n}^{(-)}\right)(2^nT)^{n+1}
+O\left(T^{n+2}\right)\right]}.
\end{align}
By comparing it with the direct Magnus expansion of $U_{n+1}^{(\pm)}=\exp\left[-i\sum_{m=0}^\infty(2^{n+1}T)^{m+1}\Omega_{n+1,m}^{(\pm)}\right]$, we obtain
\begin{equation}
\Omega_{n+1,m}^{(+)}=\Omega_{n+1,m}^{(-)}=2^{-m}\frac{\Omega_{n,m}^{(+)}+\Omega_{n,m}^{(-)}}{2}=2^{-m(n+1-m)}\frac{\Omega_{m,m}^{(+)}+\Omega_{m,m}^{(-)}}{2}
\quad \text{for all }m\leq n,
\label{eq:formula}
\end{equation}
which shows that Eq.~(\ref{eq:Magnus_formula}) also holds for $n+1$.
By induction, it is concluded that Eq.~(\ref{eq:Magnus_formula}) holds for every $n$.

In our proof, we implicitly assumed that the Magnus expansion and the expansion regarding the Baker-Campbell-Hausdorff formula are convergent. However, this assumption is justified: we introduce a sufficiently small positive constant $\epsilon>0$ and replace $H(t)$ by $\epsilon H(t)$ so that these expansions are actually convergent. This procedure does not affect the conclusion since $\{\Omega_{n,m}^{(\pm)}\}$ do not depend on $\epsilon$.

\section{Fermi's golden rule for random driving sequence}
Let us consider two Hamiltonians $H_\pm$ and assume that $H(t)$ is either $H_+$ or $H_-$ for each time period $t\in[mT,(m+1)T)$.
By introducing a stochastic variable $\xi(t)$, where $\xi(t)=+1$ or $-1$ for each period, $H(t)$ is expressed as
\begin{equation}
H(t)=\frac{H_++H_-}{2}+\xi(t)\frac{H_+-H_-}{2}=:H_0+\xi(t)V.
\end{equation}

Now let us consider the time evolution from $t=0$ to $t=MT$, where $M$ is an integer and we will finally take the limit of $M\to\infty$.
The Fourier expansion of $\xi(t)$ is given by
\begin{equation}
\xi(t)=\frac{1}{\sqrt{MT}}\sum_{n=-\infty}^\infty\xi_ne^{iz_nt}, \quad z_n=\frac{2\pi n}{MT},
\end{equation}
When $\xi(t)$ is randomly chosen from $\pm 1$ with equal probability for each period, $|\xi_n|^2$ roughly behaves as
\begin{equation}
|\xi_n|^2\sim \frac{1}{\omega}
\end{equation}
for $|z_n|\lesssim\omega$ and decays algebraically for larger $|z_n|$.

When the initial state is given by an equilibrium density matrix $e^{-\beta H_0}/Z$, where $Z=\Tr e^{-\beta H_0}$, Fermi's golden rule yields the following result:
\begin{equation}
\kappa=\frac{\pi}{N}\lim_{M\to\infty}\sum_n\sum_{a,b}z_n\delta(E_b-E_a-z_n)\frac{|\xi_n|^2}{MT}\left|\braket{E_b|V|E_a}\right|^2\frac{e^{-\beta E_a}}{Z},
\end{equation}
where $E_a$ is an eigenvalue of $H_0$ and $\ket{E_a}$ is the corresponding eigenstate.
It is noted that $\lim_{M\to\infty}\frac{1}{MT}\sum_n=\frac{1}{2\pi}\int_{-\infty}^{\infty}dz$ with $z_n\to z$.
By using $|\xi_n|^2\sim 1/\omega$, we obtain
\begin{equation}
\kappa\sim\frac{1}{2\omega}\int_{-\infty}^\infty z\sum_{a,b}\delta(E_b-E_a-z)\frac{|\braket{E_b|V|E_a}|^2}{N}\frac{e^{-\beta E_a}}{Z}.
\label{eq:heating_Fermi}
\end{equation}
We find $\kappa\propto\omega^{-1}$.
Usually, the integral in Eq.~(\ref{eq:heating_Fermi}) is finite in the thermodynamic limit.

In the main text, Fermi's golden rule is applied to the random effective Hamiltonian for $n$-RMD: $H_\mathrm{eff}^{(n,\pm)}=H_\mathrm{eff}^{(n)}+(2^nT)^n\Omega_{n,n}^{(\pm)}$.
In this case, 
\begin{equation}
V=\frac{H_\mathrm{eff}^{(n,+)}-H_\mathrm{eff}^{(n,-)}}{2}=(2^nT)^n\frac{\Omega_{n,n}^{(+)}-\Omega_{n,n}^{(-)}}{2}\propto\omega^{-n},
\end{equation}
and hence Eq.~(\ref{eq:heating_Fermi}) implies
\begin{equation}
\kappa\sim\omega^{-1}\cdot\omega^{-2n}=\omega^{-(2n+1)}.
\end{equation}
This frequency dependence of the heating rate has been numerically verified in Ref.~\cite{Zhao2020}.

\section{Formal expression and proof of the inequality (5)}

Let us consider generic quantum lattice systems.
The set of sites is denoted by $\Lambda=\{1,2,\dots,N\}$, where each $i\in\Lambda$ is an index of each site.
For a set $X\subset\Lambda$, we define $|X|$ as the number of elements in $X$.

An operator $A$ is said to be \textit{$k$-local} if it is expressed in the form
\begin{equation}
A=\sum_{X\subset\Lambda: |X|\leq k}\mathsf{a}_X,
\label{eq:k-local}
\end{equation}
where $\mathsf{a}_X$ is an operator acting nontrivially onto the set of sites $X$.
The condition $|X|\leq k$ implies that each $\mathsf{a}_X$ acts nontrivially onto at most $k$ sites.
Furthermore, if for any $i\in\Lambda$, $\{\mathsf{a}_X\}$ in Eq.~(\ref{eq:k-local}) satisfies
\begin{equation}
\sum_{X:i\in X}\|\mathsf{a}_X\|\leq g,
\label{eq:-extensive}
\end{equation}
we say that $A$ is \textit{$g$-extensive}.
It is noted that 
\begin{equation}
\|A\|\leq\sum_{X\in\Lambda: |X|\leq k}\|\mathsf{a}_X\|\leq\sum_{i=1}^N\sum_{X:i\in X}\|\mathsf{a}_X\|\leq Ng.
\end{equation}

When each $A_l$ ($l=1,2,\dots m)$ is a $k_l$-local and $g_l$-extensive operator, for any $k_B$-local operator $B=\sum_{X:|X|\leq k_B}\mathsf{b}_X$, we have
\begin{equation}
\left\|[A_m,[A_{m-1},\dots[A_1,B]\dots]]\right\|\leq\bar{B}\prod_{l=1}^m\left(2g_lK_l\right),
\label{eq:comm}
\end{equation}
where
\begin{equation}
K_l=k_B+\sum_{j=1}^{l-1}k_j \quad\text{and}\quad
\bar{B}=\sum_{X:|X|\leq k}\|\mathsf{b}_X\|.
\end{equation}

Let us assume that the Hamiltonian $H(t)$ is $k$-local and $g$-extensive for all $t$:
\begin{equation}
H(t)=\sum_{X\subset\Lambda:|X|\leq k}\mathsf{h}_X(t), \quad \sum_{X:i\in X}\|\mathsf{h}_X(t)\|\leq g.
\end{equation}
For $H(t)$ with $t\in[0,\tau]$, we consider the time evolution operator 
\begin{equation}
U_\tau=\mathcal{T}e^{-i\int_0^\tau dt\, H(t)}=:e^{-i\tilde{H}\tau}.
\end{equation}
The $n$th order truncation of the Magnus expansion of $\tilde{H}$ is denoted by
\begin{equation}
\tilde{H}^{(n)}=\sum_{l=0}^n\tau^l\Omega_l.
\label{eq:tilde_H}
\end{equation}
In Ref.~\cite{Mori2016b}, it was proved that $\Omega_l$ is $(l+1)k$-local and 
\begin{equation}
\|\Omega_l\|\leq\bar{\Omega}_l\leq\frac{2gN}{(l+1)^2}(2gk)^ll!.
\label{eq:Omega_norm}
\end{equation}

By using the useful inequality~(\ref{eq:comm}) and (\ref{eq:Omega_norm}), we can prove the following theorem.
\begin{theorem}\label{theorem}
Let us assume that $H(t)$ is $k$-local and $g$-extensive.
Then, for any $\tau$ and $n$ satisfying $4gk(n+1)\tau\leq 1/2$,
\begin{equation}
\left\|U_\tau^\dagger OU_\tau - e^{i\tilde{H}^{(n)}\tau}Oe^{-i\tilde{H}^{(n)}\tau}\right\|\leq C(n+1)!(4gk\tau)^{n+2}2^p\bar{O}
\label{eq:theorem}
\end{equation}
holds for any $pk$-local operator $O=\sum_{X\subset\Lambda:|X|\leq pk}\mathsf{o}_X$, where $p\in\mathbb{N}$, $\bar{O}=\sum_X\|\mathsf{o}_X\|$ and $C=2e+5/2$ is a numerical factor.
\end{theorem}
We will give the proof later.

\bigskip
Now we apply Theorem~\ref{theorem} to evaluate $\left\|U_\tau^\dagger \tilde{H}^{(n)}U_\tau-\tilde{H}\right\|$.
By using Eq.~(\ref{eq:tilde_H}), we obtain
\begin{align}
\left\|U_\tau^\dagger \tilde{H}^{(n)}U_\tau-\tilde{H}^{(n)}\right\|&=\left\|U_\tau^\dagger \tilde{H}^{(n)}U_\tau-e^{i\tilde{H}^{(n)}\tau}\tilde{H}^{(n)}e^{-i\tilde{H}^{(n)}\tau}\right\|
\nonumber \\
&=\left\|\sum_{l=0}^n\tau^l\left(U_\tau^\dagger\Omega_lU_\tau-e^{i\tilde{H}^{(n)}\tau}\Omega_le^{-i\tilde{H}^{(n)}\tau}\right)\right\|
\nonumber \\
&\leq \sum_{l=0}^n\tau^l\left\|U_\tau^\dagger \Omega_lU_\tau-e^{i\tilde{H}^{(n)}\tau}\Omega_le^{-i\tilde{H}^{(n)}\tau}\right\|.
\end{align}
Since $\Omega_l$ is a $(l+1)k$-local operator, Theorem~\ref{theorem} is applied by putting $O=\Omega_l$ and $p=l+1$.
By using Eq.~(\ref{eq:Omega_norm}), we obtain
\begin{equation}
\left\|U_\tau^\dagger \tilde{H}^{(n)}U_\tau-\tilde{H}^{(n)}\right\|\leq C\sum_{l=0}^n(n+1)!(4gk\tau)^{n+2}\frac{4gN}{(l+1)^2}(4gk\tau)^ll!.
\end{equation}
Because $(4gk\tau)^ll!\leq (4gk\tau l)^l\leq (4gk(n+1)\tau)^l\leq 2^{-l}$ for $l\leq n$, where we have used the assumption in Theorem~\ref{theorem}, i.e., $4gk(n+1)\tau\leq 1/2$, we finally obtain
\begin{align}
\left\|U_\tau^\dagger \tilde{H}^{(n)}U_\tau-\tilde{H}^{(n)}\right\|&\leq 4C(n+1)!(4gk\tau)^{n+2}gN\sum_{l=0}^n\frac{2^{-l}}{(l+1)^2}
\nonumber \\
&\leq 4C(n+1)!(4gk\tau)^{n+2}gN\sum_{l=0}^\infty 2^{-l}
\nonumber \\
&=8C(n+1)!(4gk\tau)^{n+2}gN.
\end{align}
This is a precise expression of Eq. (5) in the main text.

\bigskip
\noindent
\textit{Proof of Theorem~\ref{theorem}}

Following Ref.~\cite{Mori2016b}, we shall prove Theorem~\ref{theorem}.
First, we consider the Dyson expansion of $U_\tau^\dagger OU_\tau$:
\begin{equation}
U_\tau^\dagger OU_\tau=\sum_{l=0}^\infty\tau^l\mathcal{A}_lO, \quad
\mathcal{A}_lO=\frac{i^l}{\tau^l}\int_0^\tau dt_1\int_0^{t_1}dt_2\dots\int_0^{t_{l-1}}dt_l[H(t_l),[H(t_{l-1}),\dots,[H(t_1),O]\dots]].
\end{equation}
Next, let us expand $e^{i\tilde{H}^{(n)}\tau}Oe^{-i\tilde{H}^{(n)}\tau}$ as
\begin{equation}
e^{i\tilde{H}^{(n)}\tau}Oe^{-i\tilde{H}^{(n)}\tau}=\sum_{l=0}^\infty\tau^l\tilde{A}_l^{(n)}O, \quad
\tilde{\mathcal{A}}_l^{(n)}=\sum_{r=0}^l\sum_{m_1,m_2,\dots,m_r=0}^n\frac{i^r}{r!}\chi\left(\sum_{i=1}^rm_i=l-r\right)[\Omega_{m_1},[\Omega_{m_2},\dots,[\Omega_{m_r},O]\dots]].
\label{eq:tilde_A}
\end{equation}
Here, $\chi(\cdot)$ is the indicator function defined by $\chi(\text{True})=1$ and $\chi(\text{False})=0$.

When the Magnus expansion is truncated at $n$th order, $\tilde{A}_l^{(n)}$ is exact up to $(n+1)$th order with respect to $l$, i.e.,
\begin{equation}
\mathcal{A}_l=\tilde{\mathcal{A}}_l^{(n)} \quad\text{for all }l\leq n+1
\label{eq:A_identity}
\end{equation}
Equation~(\ref{eq:A_identity}) is also understood as follows.
We introduce a sufficiently small positive constant $\epsilon>0$ such that the Magnus expansion becomes convergent by replacing $H(t)\to \epsilon H(t)$.
It is always possible because the convergence of the Magnus expansion is guaranteed when $\int_0^\tau dt\,\|H(t)\|\leq\xi$, where $\xi$ is a universal constant~\cite{Blanes2009}.
By this replacement, $U_\tau$ and $e^{-i\tilde{H}^{(n)}\tau}$ become, respectively,
\begin{equation}
U_\tau=e^{-i\sum_{l=0}^\infty(\epsilon\tau)^{l+1}\Omega_l}, \quad e^{-i\tilde{H}^{(n)}\tau}=e^{-i\sum_{l=0}^n(\epsilon\tau)^{l+1}\Omega_l}.
\end{equation}
These two coincide with each other up to $(n+1)$th order in $\epsilon$.
Since $\mathcal{A}_l$ and $\tilde{\mathcal{A}}_l^{(n)}$ are independent of $\epsilon$, we can conclude Eq.~(\ref{eq:A_identity}).

Now let us evaluate the left-hand side of Eq.~(\ref{eq:theorem}):
\begin{align}
\left\|U_\tau^\dagger OU_\tau - e^{i\tilde{H}^{(n)}\tau}Oe^{-i\tilde{H}^{(n)}\tau}\right\|
&=\left\|\sum_{l=0}^\infty\tau^l\left(\mathcal{A}_lO-\tilde{\mathcal{A}}_l^{(n)}O\right)\right\|
\nonumber \\
&=\left\|\sum_{l=n+2}^\infty\tau^l\left(\mathcal{A}_lO-\tilde{\mathcal{A}}_l^{(n)}O\right)\right\|
\nonumber \\
&\leq\sum_{l=n+2}^\infty\tau^l\left\|\mathcal{A}_lO-\tilde{\mathcal{A}}_l^{(n)}O\right\|
\nonumber \\
&\leq\sum_{l=n+2}^\infty\tau^l\left(\|\mathcal{A}_lO\|+\|\tilde{\mathcal{A}}_l^{(n)}O\|\right).
\label{eq:theorem_proof}
\end{align}

First, we consider $\|\mathcal{A}_lO\|$.
We have
\begin{equation}
\|\mathcal{A}_lO\|\leq\frac{1}{\tau^l}\int_0^\tau dt_1\int_0^{t_1}dt_2\dots\int_0^{t_{l-1}}dt_l\|[H(t_l),[H(t_{l-1}),\dots,[H(t_1),O]\dots]]\|.
\end{equation}
By using Eq.~(\ref{eq:comm}), we obtain
\begin{equation}
\|\mathcal{A}_lO\|\leq\bar{O}(2gk)^l\frac{(p+l-1)!}{l!(p-1)!}\leq\bar{O}2^{p-1}(4gk)^l,
\end{equation}
where we have used $(p+l-1)!/[l!(p-1)!]\leq 2^{p+l-1}$.
By using it, we obtain
\begin{equation}
\sum_{l=n+2}^\infty\tau^l\|\mathcal{A}_lO\|\leq\sum_{l=n+2}^\infty 2^{p-1}\bar{O}(4gk\tau)^l
=(4gk\tau)^{n+2}2^{p-1}\bar{O}\sum_{l=0}^\infty(4gk\tau)^l.
\end{equation}
Since $4gk\tau\leq 4gk(n+1)\tau\leq 1/2$, the following upper bound is obtained:
\begin{equation}
\sum_{l=n+2}^\infty\tau^l\|\mathcal{A}_lO\|\leq (4gk\tau)^{n+2}2^p\bar{O}.
\label{eq:A_result}
\end{equation}

Next, we consider $\|\tilde{\mathcal{A}}_l^{(n)}O\|$.
From Eq.~(\ref{eq:tilde_A}), we have, for $l\geq 1$,
\begin{equation}
\|\tilde{\mathcal{A}}_l^{(n)}O\|\leq\sum_{r=1}^l\sum_{m_1,m_2,\dots,m_r=0}^n\frac{1}{r!}\chi\left(\sum_{i=1}^rm_i=l-r\right)\left\|[\Omega_{m_1},[\Omega_{m_2},\dots,[\Omega_{m_r},O]\dots]]\right\|.
\label{eq:tilde_A_bound}
\end{equation}
It is shown in Ref.~\cite{Mori2016b} that $\Omega_l$ is $(l+1)k$-local and $g_l$-extensive with
\begin{equation}
g_l=\frac{(2gk)^ll!}{l+1}g.
\end{equation}
By using Eq.~(\ref{eq:comm}), we obtain, under the restriction $\sum_{i=1}^rm_i=l-r$,
\begin{align}
\left\|[\Omega_{m_1},[\Omega_{m_2},\dots,[\Omega_{m_r},O]\dots]]\right\|
&\leq \bar{O}\prod_{i=1}^r\left[2g_{m_i}k\left(p+i+\sum_{j=r-i+1}^rm_j\right)\right]
\nonumber \\
&\leq\bar{O}\prod_{i=1}^r\left[2g_{m_i}k(p+l-r+i)\right]
\nonumber \\
&\leq\bar{O}(2gk)^l\frac{(p+l)!}{(p+l-r)!}m_1!m_2!\dots m_r!.
\label{eq:Omega_comm_bound}
\end{align}
Let us define
\begin{equation}
F_{n,l,r}:=\max_{m_1,\dots,m_r\in\{0,1,\dots,n\}}m_1!m_2!\dots m_r!\chi\left(\sum_{i=1}^rm_i=l-r\right).
\label{eq:F}
\end{equation}
Without loss of generality, we can assume $m_1\geq m_2\geq\dots\geq m_r$.
When $l-r\leq n$, the maximum is realized when $m_1=l-r$ and $m_2=m_3=\dots=m_r=0$.
Therefore,
\begin{equation}
F_{n,l,r}=(l-r)! \quad \text{when }l-r\leq n.
\label{eq:F1}
\end{equation}
On the other hand, when $l-r\geq n$, the maximum is realized when $m_1=n$, and
\begin{equation}
F_{n,l,r}=n!\max_{m_2,m_3,\dots,m_r\in\{0,1,\dots,n\}}m_2!m_3!\dots m_r!\chi\left(\sum_{i=2}^rm_i=l-n-r\right)=n!F_{n,l-n+1,r-1}.
\end{equation}
By using $m_i!\leq m_i^{m_i}\leq n^{m_i}$, we have $F_{n,l-n+1,r-1}\leq n^{l-r-n}$ and therefore
\begin{equation}
F_{n,l,r}\leq n!n^{l-r-n} \quad \text{when }l-r\geq n.
\label{eq:F2}
\end{equation}
By substituting Eqs.~(\ref{eq:Omega_comm_bound}) and (\ref{eq:F}) into Eq.~(\ref{eq:tilde_A_bound}), we obtain
\begin{equation}
\|\tilde{\mathcal{A}}_l^{(n)}O\|\leq\sum_{r=1}^l\sum_{m_1,m_2,\dots,m_r=0}^n\chi\left(\sum_{i=1}^rm_i=l-r\right)\frac{(p+l)!}{r!(p+l-r)!}F_{n,l,r}(2gk)^l\bar{O}.
\end{equation}
By using $(p+l)!/[r!(p+l-r)!]\leq 2^{p+l}$ and 
\begin{equation}
\sum_{m_1,m_2,\dots,m_r=0}^n\chi\left(\sum_{i=1}^rm_i=l-r\right)
\leq \sum_{m_1,m_2,\dots,m_r=0}^\infty\chi\left(\sum_{i=1}^rm_i=l-r\right)
=\frac{(l-1)!}{(l-r)!(r-1)!},
\end{equation}
as well as Eqs.~(\ref{eq:F1}) and (\ref{eq:F2}), we have
\begin{align}
\|\tilde{\mathcal{A}}_l^{(n)}O\|&\leq (4gk)^l2^p\bar{O}\sum_{r=1}^l\frac{(l-1)!}{(l-r)!(r-1)!}F_{n,l,r}
\nonumber \\
&\leq (4gk)^l2^p\bar{O}\left[\sum_{r=1}^{l-n}\frac{(l-1)!}{(l-r)!(r-1)!}n!n^{l-r-n}+\sum_{r=l-n+1}^l\frac{(l-1)!}{(r-1)!}\right]
\label{eq:tilde_A_proof}
\end{align}
By substituting
\begin{equation}
\sum_{r=1}^l\frac{(l-1)!}{(l-r)!(r-1)!}n^{l-r}=(n+1)^{l-1}
\end{equation}
and
\begin{equation}
\sum_{r=l-n+1}^l\frac{(l-1)!}{(r-1)!}\leq(l-1)!\sum_{r=0}^\infty\frac{1}{(l-n+r)!}
\leq\frac{(l-1)!}{(l-n)!}\sum_{r=0}^\infty\frac{1}{(l-n+1)^r}=\frac{(l-1)!}{(l-n)!}\frac{l-n+1}{l-n}\leq\frac{3}{2}\frac{(l-1)!}{(l-n)!}
\end{equation}
into Eq.~(\ref{eq:tilde_A_proof}), we obtain
\begin{equation}
\|\tilde{\mathcal{A}}_l^{(n)}O\|\leq(4gk)^ln^p\bar{O}n!\left[n^{-n}(n+1)^{l-1}+\frac{3}{2}\frac{(l-1)!}{n!(l-n)!}\right].
\end{equation}
 By using this bound, we have
 \begin{equation}
 \sum_{l=n+2}^\infty\tau^l\|\tilde{\mathcal{A}}_l^{(n)}O\|
 \leq 2^p\bar{O}n!(4gk\tau)^{n+2}\left[(n+1)\left(1+\frac{1}{n}\right)^n\sum_{l=0}^\infty[4gk(n+1)\tau]^l
 +\frac{3}{2}\sum_{l=0}^\infty\frac{(n+1+l)!}{n!(2+l)!}(4gk\tau)^l\right].
 \end{equation}
 By using $(1+1/n)^n\leq e$, $\sum_{l=0}^\infty[4gk(n+1)\tau]^l\leq\sum_{l=0}^\infty(1/2)^l=2$, and
 \begin{align}
 \sum_{l=0}^\infty\frac{(n+1+l)!}{n!(2+l)!}(4gk\tau)^l
 &\leq\frac{1}{2}\sum_{l=0}^\infty\frac{n+1}{1}\frac{n+2}{2}\dots\frac{n+l+1}{l+1}(4gk\tau)^l
 \nonumber \\
 &\leq\frac{1}{2}\sum_{l=0}^\infty(n+1)^{l+1}(4gk\tau)^l
 \nonumber \\
 &\leq\frac{n+1}{2}\sum_{l=0}^\infty2^{-l}=n+1,
 \end{align}
 we obtain
 \begin{equation}
 \sum_{l=n+2}^\infty\tau^l\|\tilde{\mathcal{A}}_l^{(n)}O\|\leq\left(2e+\frac{3}{2}\right)(n+1)!(4gk\tau)^{n+2}2^p\bar{O}.
 \label{eq:tilde_A_result}
 \end{equation}
 
 By substituting Eqs.~(\ref{eq:A_result}) and (\ref{eq:tilde_A_result}) into Eq.~(\ref{eq:theorem_proof}), we finally obtain the desired inequality:
 \begin{equation}
 \left\|U_\tau^\dagger OU_\tau-e^{i\tilde{H}^{(n)}\tau}Oe^{-i\tilde{H}^{(n)}\tau}\right\|
 \leq(4gk\tau)^{n+2}2^p\bar{O}+\left(2e+\frac{3}{2}\right)(n+1)!(4gk\tau)^{n+2}2^p\bar{O}
 \leq \left(2e+\frac{5}{2}\right)(n+1)!(4gk\tau)^{n+2}2^p\bar{O}.
 \end{equation}

\bibliography{apsrevcontrol,physics}